\documentclass[prb, twocolumn, amsmath, amsfonts, amssymb, amsthm]{revtex4-2}

\usepackage[utf8]{inputenc}
\usepackage[T1]{fontenc}
\usepackage[main=english]{babel}

\usepackage{hyperref}
\hypersetup{colorlinks=true, linkcolor=blue, citecolor=blue, urlcolor=blue, filecolor=red, pdfstartview=}

\newcommand*{\iu}{\mathrm{i}}
\newcommand*{\Elr}{\mathrm{e}}
\newcommand*{\vb}[1]{\mathbf{#1}}
\newcommand*{\ev}[1]{\langle {#1} \rangle}
\DeclareMathOperator{\tr}{tr}

\begin{document}
\title{Comment on ``Towards exact solutions for the superconducting \boldmath{$T_c$} induced by electron-phonon interaction''}
\author{Grgur Palle}
\affiliation{Institute for Theoretical Condensed Matter Physics, Karlsruhe Institute of Technology, 76131 Karlsruhe, Germany}
\date{\today}
\begin{abstract}
In a series of recent articles, Liu, Yang, Pan, \textit{et al.}\ claim to have determined the exact dressed electron-boson vertex for a number of very general interacting many-body problems by inverting certain Ward-Takahashi identities.
Here, we point out that their Ward-Takahashi identities are missing terms which make the inversion of the identities impossible.
One therefore cannot formulate exact self-contained integral equations for the dressed electron propagator.
In addition, the proposed vertex expressions do not reproduce well-established results for the leading perturbative corrections.
\end{abstract}

\maketitle

One of the outstanding problems in treating strong electron-phonon interactions beyond Migdal's approximation is calculating vertex corrections.
To date, most approaches to this problem have been approximate, with crucial constraints on the vertex corrections coming from conservation laws in the form of Ward-Takahashi identities (WTI).
By using Heisenberg's equations of motion, one may relate the dressed vertex exactly to higher-order Green functions, whose equations of motion are then related to even higher-order Green functions.
The problem of calculating the dressed electron-phonon vertex may thus be reformulated as the problem of truncating this infinite hierarchy of equations of motion.
It is widely believed that there are no easy solutions to this problem in higher dimensions, especially not exact ones, valid for generic values of the interaction strength.

It thus comes as a surprise to read~\cite{LYPW} that the charge and spin WTIs are enough to express the dressed electron-phonon vertex exactly in terms of the dressed electron propagator.
If true, such a relation would enable one to formulate a self-contained integral equation for the electron propagator that yields the exact solution of the interacting electron-phonon problem, but is no more difficult than self-consistent Hartree-Fock.
Generic density-density interactions would become solvable as well, not only those mediated by phonons, but also strong electron-electron Coulomb interactions.
In subsequent work~\cite{LYPH, LYP-Dirac1, LYP-Dirac2, LYP-Dirac3}, the same authors indeed claim that WTIs can be inverted in other models with density-density interactions to give exact self-contained integral equations for the fermionic propagators.
In this manner, Liu, Yang, Pan, \textit{et al.}\ have studied interacting Dirac fermion systems~\cite{LYP-Dirac1}, the interplay between Coulomb and electron-phonon coupling~\cite{LYPH, LYP-Dirac2}, and related problems~\cite{LYP-Dirac3}.

\begin{widetext}
In this Comment, we show that the proposed~\cite{LYPW} relation between the dressed vertex and electron propagator does not hold.
This is most directly seen by comparing its predictions with perturbation theory.
For the model considered in Ref.~\cite{LYPW}, one readily finds the electron-density vertex that enters the charge WTIs:
\begin{equation*}
\Gamma_t(q, p) = \sigma_3 + \iu g^2 \int_k G_0(p+q+k) \sigma_3 G_0(p+k) D_0(k) - \iu g^2 \sigma_3 \, D_0(q) \int_k \tr \sigma_3 G_0(p+q+k) \sigma_3 G_0(p+k) + \mathcal{O}(g^4),
\end{equation*}
where $G_0(p) = (\omega_p \sigma_0 - \xi_{\vb{p}} \sigma_3)^{-1}$ and $D_0(q) = 2 \Omega_{\vb{q}} / (\omega_q^2 - \Omega_{\vb{q}}^2)$ are the bare electron and phonon propagators, $g$ is the electron-phonon coupling constant, and $\sigma_{\mu}$ are Pauli matrices in Nambu space.
On the other hand, if one inserts the Fock self-energy $\Sigma(p) = \iu g^2 \int_k G_0(p+k) D_0(k) + \mathcal{O}(g^4)$ into the proposed Eq.~(40) of Ref.~\cite{LYPW}, one obtains:
\begin{equation*}
\Gamma_t^{\text{Ref.\cite{LYPW}}}(q,p) = \sigma_3 + \iu g^2 \left[G_0^{-1}(p+q) - G_0^{-1}(p)\right]^{-1} \sigma_3 \int_k \left[G_0(p+k) - G_0(p+q+k)\right] D_0(k) + \mathcal{O}(g^4).
\end{equation*}
\end{widetext}
Evidently, the proposed exact expression for $\Gamma_t$ does not reproduce well-established weak-coupling expansions and cannot be correct.

The source of the mistake is the following:
In the charge and spin WTIs, the averages which include current divergences may be expressed via limits
\begin{gather*}
\left\langle\left[\xi_{\partial_{\vb{z}}}\!\Psi^{\dag}(z) \sigma_m \Psi(z) - \Psi^{\dag}(z) \sigma_m \xi_{\partial_{\vb{z}}}\!\Psi(z)\right] \Psi(z_1) \Psi^{\dag}(z_2)\right\rangle \\
= \lim_{z' \to z} \left(\xi_{\partial_{\vb{z}'}} - \xi_{\partial_{\vb{z}}}\right) \ev{\Psi^{\dag}(z') \sigma_m \Psi(z) \Psi(z_1) \Psi^{\dag}(z_2)}
\end{gather*}
to obtain the Eqs.~(35) and~(36) of Ref.~\cite{LYPW}, which are correct.
However, in going to Eqs.~(37) and~(38), which are incorrect, the dependence of the above averages on the difference $z-z'$ has been neglected.
Even though the $z' \to z$ limit is taken at the end, the dependence on $z-z'$ must be taken into account during the intermediate steps to be able to describe $z$ derivatives~\cite{fnt}.
To take this dependence into account, write
\begin{gather*}
\ev{\Psi^{\dag}(z') \sigma_m \Psi(z) \Psi(z_1) \Psi^{\dag}(z_2)} = \\
= - \int dz_3 dz_4 G(z_1-z_3) \tilde{\Gamma}_m(z', z, z_3, z_4) G(z_4-z_2), \\
\tilde{\Gamma}_m(z', z, z_3, z_4) = \tilde{\Gamma}_m(z_3 - z', z - z_4, z-z') \\
= \int_{q,p,k} \Elr^{- \iu (p+q) (z_3-z')} \Elr^{- \iu p (z-z_4)} \Elr^{- \iu k (z-z')} \tilde{\Gamma}_m(q,p,k),
\end{gather*}
where $m = t$ corresponds to $\sigma_m = \sigma_3$ (charge), and $m = s$ to $\sigma_m = \sigma_0$ (spin).
Note that $\Gamma_m(q, p) = \int_k \tilde{\Gamma}_m(q, p, k)$.

From Eqs.~(35) and~(36) of Ref.~\cite{LYPW}, instead of Eqs.~(37) and~(38), we now obtain the WTIs:
\begin{gather*}
\begin{gathered}
\omega_q \Gamma_m(q, p) - (\xi_{\vb{p}+\vb{q}} - \xi_{\vb{p}}) \Gamma_{\overline{m}}(q, p) - \Delta_{\overline{m}} = \\
= G^{-1}(p+q) \sigma_m - \sigma_m G^{-1}(p),
\end{gathered} \label{eq:my-WTI}
\end{gather*}
where $\bar{t} = s$ and $\bar{s} = t$, so $\sigma_{\overline{m}} = \sigma_m \sigma_3$. The additional
\begin{align*}
\Delta_m &= \int_k \left[\xi_{\vb{p}+\vb{q}+\vb{k}} - \xi_{\vb{p}+\vb{k}} - \xi_{\vb{p}+\vb{q}} + \xi_{\vb{p}}\right] \tilde{\Gamma}_m(q,p,k)
\end{align*}
terms make it impossible to solve these WTIs for $\Gamma_m(q, p)$ purely in terms of the dressed electron propagator $G(p)$.

The $\Delta_m$ terms are essential in ensuring that the WTI holds at each order in perturbation theory.
This is most easily seen if we recast it into
\begin{gather*}
\int_k \left[(\omega_p+\omega_q+\omega_k) - (\omega_p+\omega_k)\right] \tilde{\Gamma}_m^{\text{ir}}(q, p, k) \\[-4pt]
- \int_k \left[\xi_{\vb{p}+\vb{q}+\vb{k}} - \xi_{\vb{p}+\vb{k}}\right] \tilde{\Gamma}_{\overline{m}}^{\text{ir}}(q, p, k) = \\
= \sigma_m \Sigma(p) - \Sigma(p+q) \sigma_m,
\end{gather*}
where $\tilde{\Gamma}_m^{\text{ir}}$ is the irreducible part of $\tilde{\Gamma}_m$.
By inspecting this WTI, one sees that in $\tilde{\Gamma}_m^{\text{ir}}(q, p, k)$ every $\sigma_m$ is in effect replaced by $\sigma_m [G_0^{-1}(p+q+k) - G_0^{-1}(p+k)]$.
At each order in $g$, this type of cutting of irreducible diagrams of $\tilde{\Gamma}_m^{\text{ir}}$ on the left-hand side yields precisely the self-energy diagrams of $\Sigma$ on the right-hand side~\cite{Peskin}.
The perturbative expressions for $\Gamma_t$ and $\Sigma$ that we previously employed indeed satisfy this, as one may check.

Analogous mistakes were made in subsequent work by the same authors~\cite{LYPH, LYP-Dirac1, LYP-Dirac2, LYP-Dirac3}.
In none of the studied models~\cite{LYPW, LYPH, LYP-Dirac1, LYP-Dirac2, LYP-Dirac3} can one solve the WTIs to get the dressed fermion-boson vertex in terms of the dressed fermionic propagators only.
Moreover, if one neglects terms of the WTIs as in Refs.~\cite{LYPW, LYPH, LYP-Dirac1, LYP-Dirac2, LYP-Dirac3}, one obtains expressions for the fermion-boson vertex which violate conservation laws and do not reproduce well-established weak-coupling expansions.
The proposed vertex expressions are thus not good approximations of the exact vertex.

\textit{Acknowledgments:} I thank Jörg Schmalian, Rafael M.\ Fernandes, and Andrey V.\ Chubukov for useful suggestions.
Discussions with Guo-Zhu Liu are acknowledged.
This work was supported by the Deutsche Forschungsgemeinschaft (DFG, German Research Foundation) - TRR 288-422213477 Elasto-Q-Mat project A07.


\begin{thebibliography}{9}
\bibitem{fnt}
Averages of the form $\ev{\Psi^{\dag}(z) \sigma_m \partial_z \Psi(z) \Psi(z_1) \Psi^{\dag}(z_2)}$ contain information not only on $z' = z$, but also on $z'$ infinitesimally close to $z$, hence one cannot relate them to $\ev{\Psi^{\dag}(z) \sigma_m \Psi(z) \Psi(z_1) \Psi^{\dag}(z_2)}$.
\end{thebibliography}
\end{document}